# Application and research review of cemented carbide heterogeneous brazing coating additive manufacturing

Datian Cui[a,b], Sujuan Zhong[b], Kexing Song[c], Xingxing Wang[a], Xiwang Zhang[a], Weimin Long[b,d],＊

[a] School of Materials Science and Engineering, North China University of Water Resources and Electric Power, Zhengzhou 450045, China

[b] State Key Laboratory of Advanced Brazing Filler Metals & Technology, Zhengzhou Research Institute of Mechanical Engineering Co., Ltd, Zhengzhou 450001, China

[c] School of Materials Science and Engineering, Henan University of Science and Technology, Luoyang 471023, China

[d] China Innovation Academy of Intelligent Equipment (Ningbo) Co., Ltd, Ningbo 315700, China

*Abstract:* As one of the important methods for surface modification of materials and life extension of key components, cemented carbide brazing coatings are widely used in agricultural machinery, oil drilling, aerospace and other fields, it has also attracted the attention of scholars in the field of surface modification at home and abroad. Based on the research reports of recent 20 years at home and abroad, the present research situation of cemented carbide brazed coating additive manufacturing technology are reviewed firstly. The research progress in preparation and performance control of cemented carbide/iron-base, cemented carbide/copper-base, cemented carbide/nickel-base, cemented carbide/silver-base heterogeneous brazing coatings are reviewed in detail. Then the practical applications of cemented carbide heterogenous brazing coatings in the fields of contact soil agricultural machinery parts life extension, aviation parts repair, surface function strengthening and so on are reviewed. In this review, the limitation of the research of cemented carbide heterogenous braze coating technology is discussed. The deficiencies in the research and development of cemented carbide heterogenous braze coating technology are summarized including cemented carbide heterogeneous additive materials and technology need to be expanded, the structure of cemented carbide heterogeneous brazing coatings need to be optimized, the mechanism of interface defects in cemented carbide heterogeneous brazing coatings need to be clarified. Finally, the future development direction of braze coating technology is prospected, too.

## 1. Introduction

Additive manufacturing technology is a new technology that can integrate advanced manufacturing, intelligent manufacturing, green manufacturing, new materials, precision control and other technologies[1-3]. Due to its numerous advantages, additive manufacturing has become a popular trend in manufacturing processes[4,5]. Heterogeneous brazing additive manufacturing is a new breakthrough as well as new industry in additive manufacturing[6,7]. It has broad application prospects, especially in repairing and surface hardening of easily worn parts[8-10]. Cemented carbide heterogeneous brazing coating additive manufacturing is to connect hard particles with high hardness, wear resistance, corrosion resistance or oxidation resistance to the substrate surface by brazing, so as to realize the additive manufacturing of composite coating. It has the unique advantages of high surface precision, less machining process, high bonding strength, low heating temperature, low thermal stress, high wear resistance, high corrosion resistance and high temperature resistance [11−13]. So it is widely used in many fields such as aerospace, agricultural machinery, oil drilling coal mining machinery, and has also been highly concerned by scholars at home and abroad[14−16].

At present, the research on carbide composite brazing coating can be summarized as follows [17−20]: (1) Selection and optimization of high melting point cemented carbide components. Cemented carbide with high melting point is an essential component in the preparation of cemented carbide composite brazing coatings, which should meet the requirement of good wettability, low brittleness and low cost as well as high hardness. WC and $Cr_3C_2$ are the most common options at present. (2) Components optimization and modification of the filler metal. Brazing filler as an important component of brazing coatings. With high hardness, nickel base brazing filler and copper base brazing filler are the mainly common brazing filler, such as NiCrBSi brazing filler, NiCrP brazing filler, CuMnNi brazing filler and so on. (3) Optimization and innovation of the preparation method of brazing coating. Vacuum brazing, induction brazing, argon arc brazing, laser brazing and flame brazing are developed to prepare high performance composite coatings, which is made from the powder composed of high melting point carbide and low melting point brazing fillers by means of vacuum brazing, induction heating, argon arc heating, flame heating and laser heating. By combining brazing coating technology and bionics, biomimetic composite coatings with special morphology of biological surface have been designed and prepared, such as non-smooth wear-resistant, corrosion-resistant and super-hard coating with morphology of dung beetle, abalone and shark.

Heterogeneous brazing coating has a relatively short development process, and its additive manufacturing mechanism, technology, special equipment and application effect need to be systematically studied[21-23]. In view of the wide application potential of cemented carbide heterogeneous brazing coating additive manufacturing, which can meet the manufacturing needs of aerospace, agricultural machinery, engineering equipment and other important fields, the authors review the research progress of cemented carbide heterogeneous brazing additive manufacturing, and discuss the key problems existing in this paper. This paper focuses on the interface properties, process control and engineering application of cemented carbide heterogeneous brazing coating additive, and the problems and future development direction of cemented carbide heterogeneous brazing coating additive manufacturing are put forward too, which can provide reference for the innovation of new additive manufacturing methods and the expansion of new additive manufacturing applications.

## 2. Research overview of cemented carbide heterogeneous brazing coating additive manufacturing

Corresponding to the brazing method, the common brazing coating methods are mainly refer to vacuum braze coating (reactive braze coating), induction braze coating, laser braze coating, argon arc braze coating, flame braze coating, among which vacuum braze coating (vacuum reactive braze coating) is the hottest research issue and has been used widely[24]. The research results of brazing coating technology in the past twenty years are mainly focused on the preparation process of composite coating, the regulation and influence of low melting point filler metal components, cemented cemented carbide types, content and particle size on the microstructure and properties of the composite coating. According to the different types of filler metals and cemented carbides, four categories of 12 kinds of composite brazing coatings, namely Nickel /WC, Nickel /$Cr_3C_2$, Nickel /$B_4C$, Copper /TiC, copper /WC, Iron /WC, Iron /$Cr_3C_2$, Iron /TiC, Iron /CrB, Silver /WB, Silver/diamond and Silver/c-BN, have been successfully prepared by combing four kinds of filler metals including Ni-based, Cu-based, Fe-based and Ag-based fillers with seven kinds of cemented carbides (ceramics) such as WC, $Cr_3C_2$, TiC, $B_4C$, WB, diamond and c-BN.There are more than twenty domestic and foreign scientific research institutions developing brazing coating, including University of Science and Technology Beijing, Institute of Metal Research, Chinese Academy of Sciences, Jilin University, Zhengzhou Research Institute of Mechanical Engineering Co., Ltd and Tarbiat Modares University, and more than seventy research results have been published on brazing coating additive. The representative research results of major research institutions are shown in Table 1.

Table 1. The representative research results of major institutions

| Research institute | Brazing alloy / cemented carbide | Related reference |
| --- | --- | --- |
| University of Science and Technology Beijing | Fe-based/ TiC,$Cr_3C_2$ | [25-27] |
| Kunming University of Science and Technology | Fe-based /WC | [28] |
| Indian Institute of Technology Kanpur | Fe-based/CrB | [29] |
| Huazhong University of Science and Technology | Cu-based /WC | [30] |
| Jilin University | | [31-34] |
| China Academy of Machinery Science and Technology Group | | [35] |
| University of Missouri | | [36] |
| Southeast University | | [37-38] |
| Jilin University | Cu-based /TiC | [39] |
| Zhengzhou Research Institute of Mechanical Engineering Co. LTD | Ni-based /WC | [40-41] |
| Institute of Metals, Chinese Academy of Sciences | | [42-43] |
| Shanghai University of Electric Power | | [44-45] |
| USTHB | | [46] |
| Tarbiat Modares University, | | [47] |
| Sichuan University | | [48] |
| Jiangsu University of Science and Technology Tsingtao | | [49-50] |
| University of Science and Technology | | [51] |
| Nanchang Institute of Technology | | [52] |
| Jiangxi Academy of Sciences | Ni-based /$Cr_3C_2$ | [53-54] |
| Nanchang University | | [55] |
| Tiandi Science & Technology Co.,Ltd. | Ni-based /$B_4C$ | [56] |
| Hebei University of Technology | AgCu active alloy /Diamond | [57] |
| Harbin Institute of Technology | AgCu active alloy/WB | [58] |
| Jilin University | Ag-based/C-BN | [59] |

## 3. Research progress on properties and process control of cemented carbide heterogeneous brazing coating additive

In order to ensure the bonding strength and required properties of brazing coatings, iron base, copper base, nickel base and silver base brazing fillers are generally used as bonding phases. The brazing filler has the function of consolidating cemented carbide and bearing load [60], while the cemented carbide has the function of strengthening the surface properties of components. The filler and strengthened phase cemented carbide in the brazing coating interact, complement and restrict each other [61,62], and ultimately determine the service performance of cemented carbide composite brazing coating together. The following is a detailed review of the research progress in the preparation technology, microstructure and property control of brazing coatings on cemented carbide/iron base filler metal, cemented carbide/copper base filler metal, cemented carbide/nickel base filler metal and cemented carbide/silver base filler metal.

*3.1. Carbide/iron base heterogeneous brazing coating*

Compared with other brazing coatings, the material cost is low to use cheap industrial ferrotitanium and ferrochrome powder directly. As the Fe-base alloy has good compatibility with ordinary steel matrix, quite dense texture [44,45] can be obtained. In order to reduce the cost of materials and improve the stability of brazing coatings, domestic scholars prepared WC/Fe, TiC/Fe heterogeneous brazing coatings by using ferrotitanium powder, ferrochrome powder and ferroboron powder as the main raw materials. Pei et al. [25,27] successfully prepared carbide Fe-base alloy composite coating on low carbon steel matrix with vacuum reactive brazing by using titanium powder, ferrotitanium powder, iron powder, colloidal graphite and $Cr_3C_2$ powder as raw materials. Zhou et al. [28] found that the WC particles reinforced iron matrix composites can be fabricated by the vacuum infiltration casting technique, with the volume fraction of WC changed from low to high as 52 vol.%. With the increase of WC particle volume fraction, the wear rate of WC/Fe composite coating first decreased and then increased, which was 1.38~2.93 times lower than that of high chromium cast iron in general, indicating a significant improvement in wear resistance. Fig. 1(a) and (b) show the microstructures of the composite coatings containing 36 and 19 vol.% WC particles, respectively. These composites consist of well-distributed big WC particles, small carbide particles and bars, and iron matrix. A filler metal powder containing a mixture of Mo, Fe, Cr, MoB and FeB is uniformly dispersed on the surface of the metal matrix to form the interface structure of composite boride structure, and the braze coatings can be used for applications involving wear resistance like pump impeller parts, machining tools, and injection molding screws .

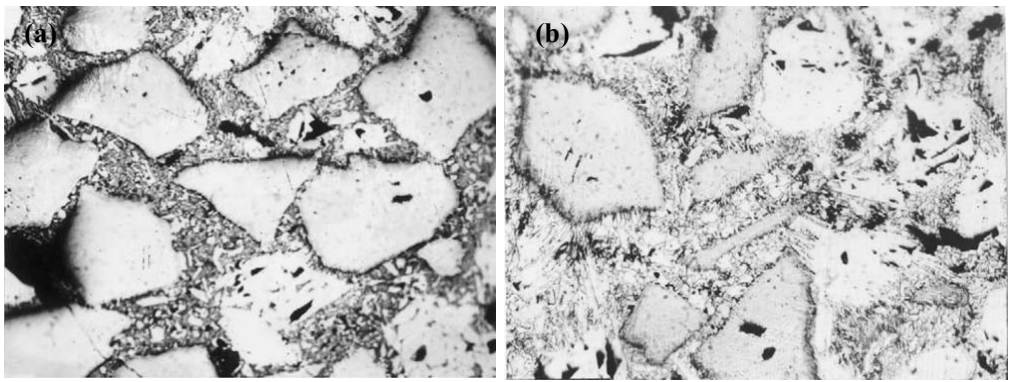

**Fig.1.** The microstructures of the composites (OP)[28].

Cemented carbide/iron base brazing coating additive manufacturing can be used in the wear resistant components such as pump impeller parts and injection molding screws. The above technology can alter the thickness of brazing coating easily, the diffusion drive between the coating and substrate has also improved the interface bonding strength, and 50%Fe dispersed in the substrate helps to reduce coating wear [18]. However, during the brazing process, the slurry coating on the surface of steel matrix has a certain fluidity, which would result in poor uniformity of the coating thickness.

*3.2. Carbide/copper base heterogeneous brazing coating*

By adding Ni, Mn, Co and other elements to optimize the composition of copper solder, the cemented carbide/copper base heterogeneous brazing coating can rapidly spread on the surface of the matrix, and the corrosion resistance and wear resistance of the brazing coating can be improved. Qi et al. [35] studied the influence of different particle size and mass fraction of tungsten carbide on the microstructure, mechanical properties, bonding strength and wear resistance of WC/CuZnNi brazing coating. It was found that WC/Cu composite coating can improve the wear resistance of Q235A steel surface. The higher the WC content, the better the wear resistance of the coatings, and the average porosity of the coating is 3.42%. Jun et al.[30] investigated the microstructure and wear resistance performance of the fabricated Cu-Ni-Mn alloy based hardfacing coating reinforced by WC particles, it is the comprehensive results of "protect function" of the WC particles, and "support function" of the Cu-Ni-Mn metal matrix that makes the WC-reinforced composite coating obtain good wear resistance performance. Fig.2 is higher-magnification SEM images of the worn surfaces. Obvious traces of ductile flow of the matrix can be observed in the worn surface of two kinds of materials caused by silica sands. The major wear mechanisms are the plastic deformation and ductile flow of the Cu-Ni-Mn matrix, and the fracturing of WC particles.

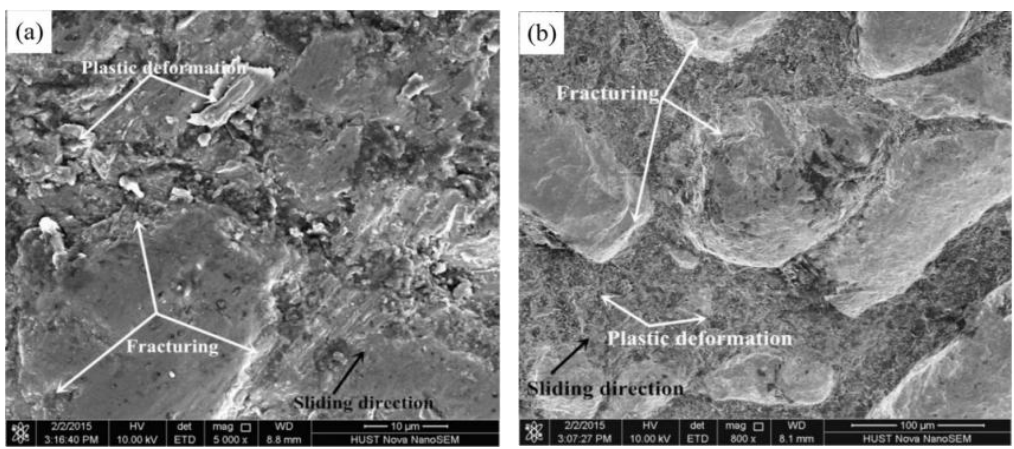

**Fig.2.** High-magnification SEM photographs of worn surfaces (a) high-Cr cast iron and (b) tungsten carbide reinforced composite[30].

The tribological performance of brazing coating is an important index which is concerned by researchers. Wear resistant composite coating SiC/Cu has been prepared on 45# steel by induction brazing process. Nickel on SiC surface can improve the wettability of copper solder to SiC. The wear mechanism of the coating is micro-cutting, furling and brittle phase fracture and shedding. Xu et al. [33,34] brazed WC particles on the surface of 45# steel by heating multiple copper base solder and brazing agent to form a bionic non-smooth wear-resistant composite coating with high-frequency induction equipment. Direct contact between WC particles and the reinforced matrix was not found in the coating. Pan et al. [37,38] successfully prepared WC-Cu composite brazing coating with Cu64MnNi as filler metal, metallurgical bonding had formed between the coating and 45# steel matrix. The wear resistance of the coating increased with the number of WC particles. When the WC particle content (mass fraction) was 30% and the average particle size was 150μm, the wear resistance of the coating was the best, which was better than that of the 45#

steel after heat treatment. But the bonding strength of composite coating decreased, leading to WC shedding in friction test, and the wear resistance of coating became worse.

WC/ Copper wear resistant brazing coating was prepared on the surface of low carbon steel by high temperature brazing process [36]. The results show that compared with the common WC-Co coating, the brazing coating has better wear resistance. Good metallurgical bonding is formed at the interface of WC/Cu alloy and composite coating/matrix, with no obvious porosity and high bonding strength. However, the coupling mechanism of cemented carbide/cu-based solder heterogeneous brazing coating remains to be further studied.

*3.3. Carbide/nickel base heterogeneous brazing coating*

The nickel-based filler metal has good corrosion resistance, high temperature resistance and oxidation resistance, and can be used for the preparation of brazing coating of relevant components [63,64] in high temperature environment. It is distributed in the form of intermetallic compounds in the brazing coating and has good corrosion resistance, high temperature resistance and oxidation resistance. It can be used for brazing components in high temperature service.

The formation mechanism of laser brazing coatings under different scanning speeds was investigated by Long et al. [2] , the Ni-based filler alloy (BNi-2 solder) was used to braze diamond particles to a steel substrate via laser in an argon . It was found that different scanning speed leaded to different heating conditions of solder layer, and finally leaded to different melting behavior of the brazing coating. The schematic diagram of formation mechanism of laser brazing diamond coating with different laser scanning speeds is shown in Fig.3. Long et al. [41] prepared wear-resisting WC-Ni coatings by flame brazing process and found that the metallurgical reaction between Ni solder and WC reinforced phase weakened with the increase of the particle size of WC. When WC particle size is 18-23 μm, long-shaped Ni-W phase appears in the interfacial transition zone of the brazing coating. The wear resistance of the brazing coating is the best when the particle size of WC is 250 ~ 380μm, microstructure of the interface between WC-Ni coating and steel is shown as Fig.4. Appropriate amount of YG8 significantly reduces the porosity of nickel-based WC− Nickel-based WC heterogeneous brazing composite coating and improves the density of the coating, thus improving the hardness and wear resistance of the coating [65].

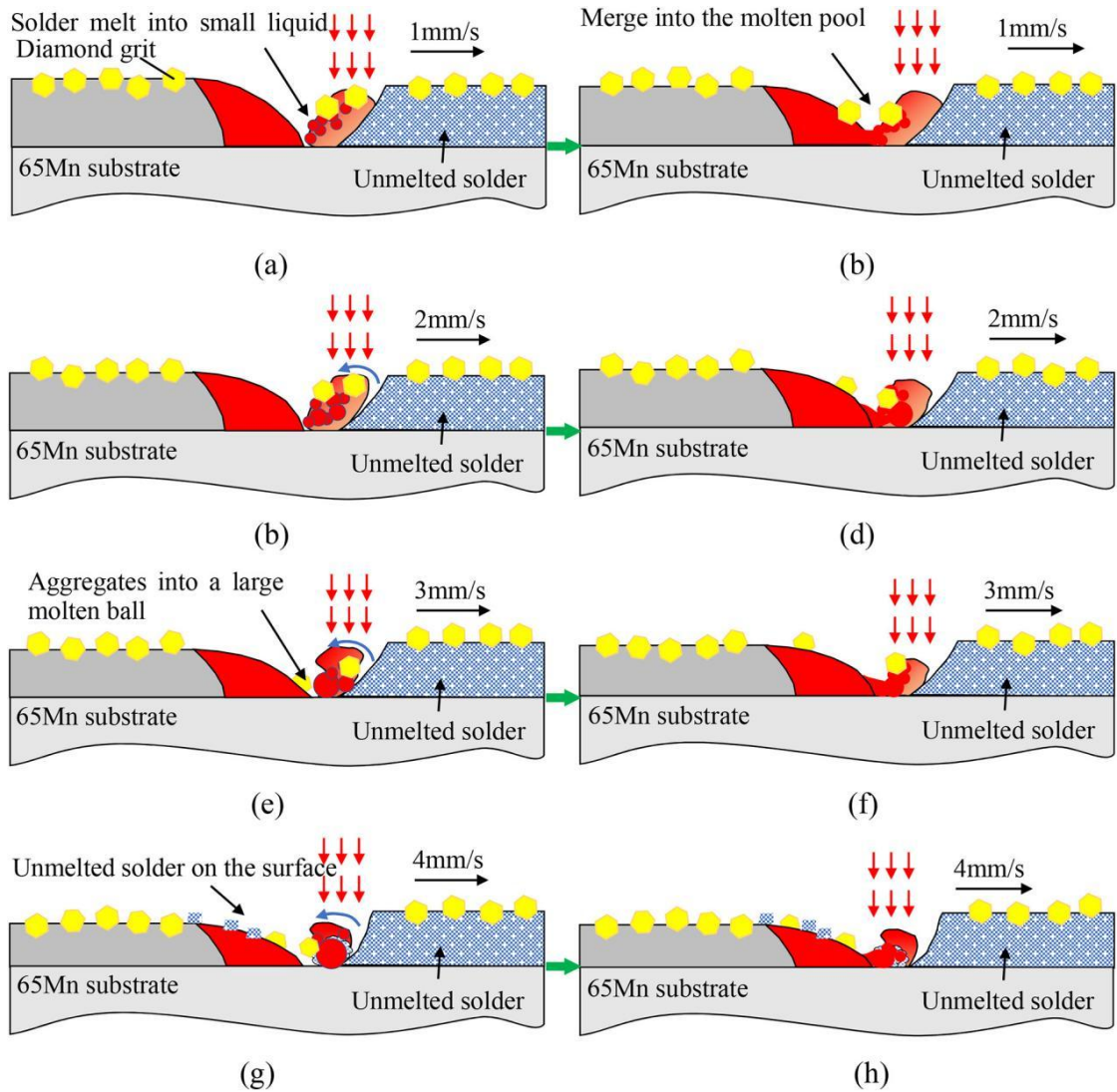

**Fig. 3.** Schematic diagram of formation mechanism of laser brazing diamond coating with different laser scanning speeds, (a and b) 1 mm/s, (c and d) 2 mm/s, (e and f) 3 mm/s, (g and h) 4 mm/s[2].

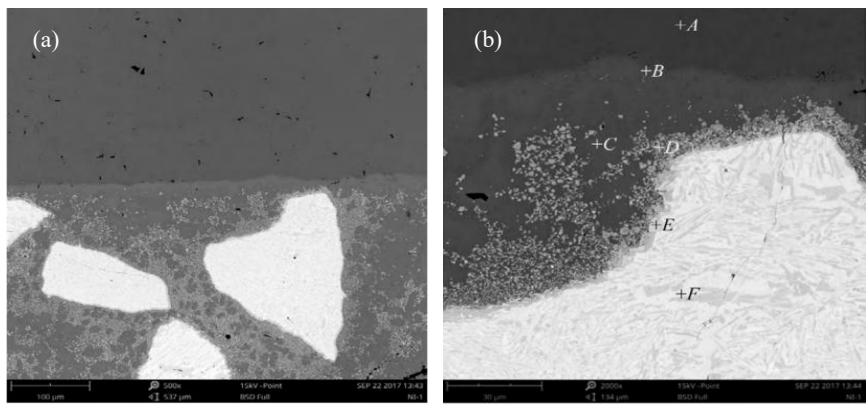

**Fig.4.** Microstructure of interface between WC-Ni coating and steel [41](a)Low magnification; (b)High magnification.

Xuan et al. [66] have prepared gradient WC/NiCrBSi composite coating by vacuum brazing. As the gradient slope of the coating increases, the wear amount of the coating decreases and wear resistance increases. After wearing 5000 minutes, the wear amount of the stainless steel matrix is 25 ~ 70 times that of the coating, and the wear resistance of the matrix is improved by the gradient coating significantly. However, the tensile strength decreases with the increase of coating gradient slope. Flexible cloth brazing coating technology is a new surface modification coating technology [49]. Lu et al. [43,67−69] prepared flexible metal cloth by mixing WC powder, NiCrBSi powder and organic adhesive with rolling process successfully. The metal cloth was prefabricated on the surface of the parts, and the wear resistant brazing coating was made by high temperature brazing. The bonding strength of the coating increased with the increase of the content of WC-17Co. The bonding strength was up to 140MPa, and the bonding strength between the coating and the matrix was 360MPa.

Prefabricated coating on cemented carbide surface can improve the bonding properties of the cemented carbide heterogeneous brazing coating. The copper shell with a thickness between 30 nm and 50nm has been prepared on the surface of ultrafine WC particles by electroless plating, and the microstructure of the brazing coating is uniform ultrafine crystal compared with that of WC-free fund ceramic cladding. The hardness of the new-type composite coating is almost 1500HV, which is 40% higher than that of the copper-free coatings. The Cu coating can inhibit the abnormal growth of WC grains and improve the crack initiation resistance of the coating significantly [47]. Feng et al. [48] found that the shear strength of WC-Co/Nickel layer /3Cr13 joint first increases and then decreases with the increase of brazing temperature and the extension of brazing time. At 1100℃ for 10min, the maximum shear strength of the brazing joint is 154MPa. Fractures of the brazing joint occur on the WC-CO matrix next to the middle layer.

The content and type of cemented carbide have a significant effect on the comprehensive properties of brazing coating. Wang et al. [55] used laser cladding (LC) to fabricate NiCoCrAlY–$B_4C$ composite coatings with $B_4C$ of 5%, 10% and 15% (mass fraction) . The effects of $B_4C$ mass fraction on the coefficient of friction (COF) and wear rate of NiCoCrAlY–$B_4C$ coatings were investigated. Morphologies of worn tracks on Ni-based coatings with different $B_4C$ mass fractions are shown as Fig.5.

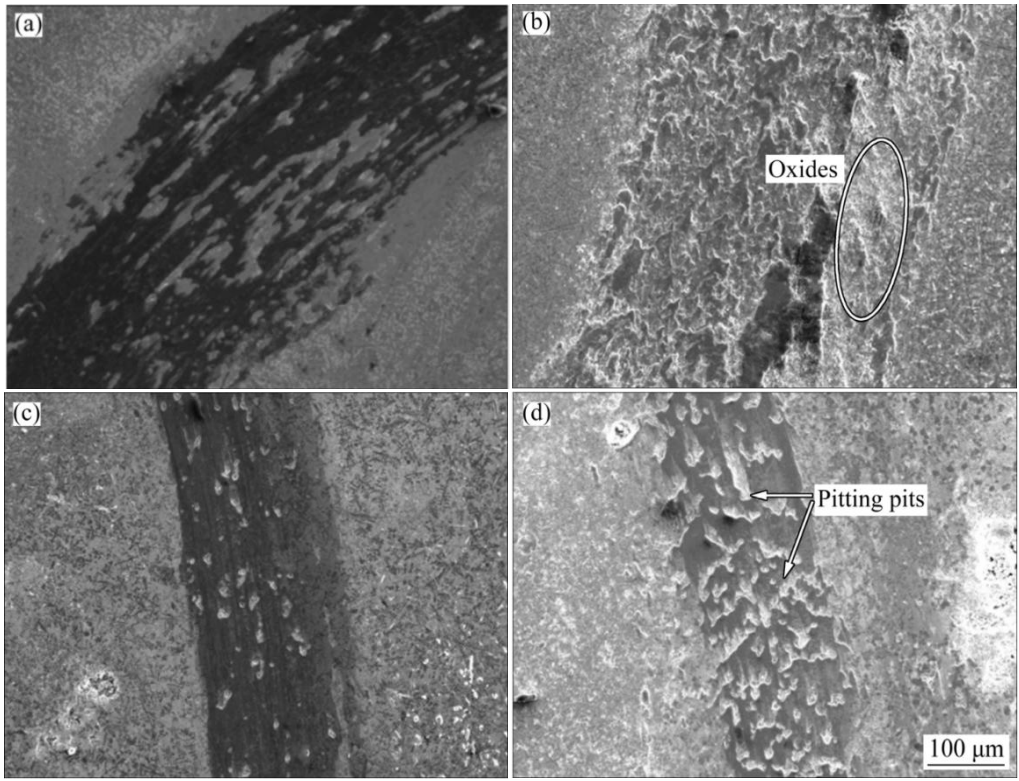

**Fig.5.** Morphologies of worn tracks on NiCoCrAlY coatings with B$_4$C mass fractions of 0 (a), 5% (b), 10% (c) and15% (d)[55].

The results show that the COFs and wear rates of Ni-based/B$_4$C coatings decrease with the increase of B$_4$C content, which are contributed to the improvement of coating hardness by the B$_4$C addition. The wear mechanisms of Ni-based/B$_4$C coatings are changed from adhesive wear and oxidation wear to fatigue wear with the increase of B$_4$C content, as shown in Fig.6.

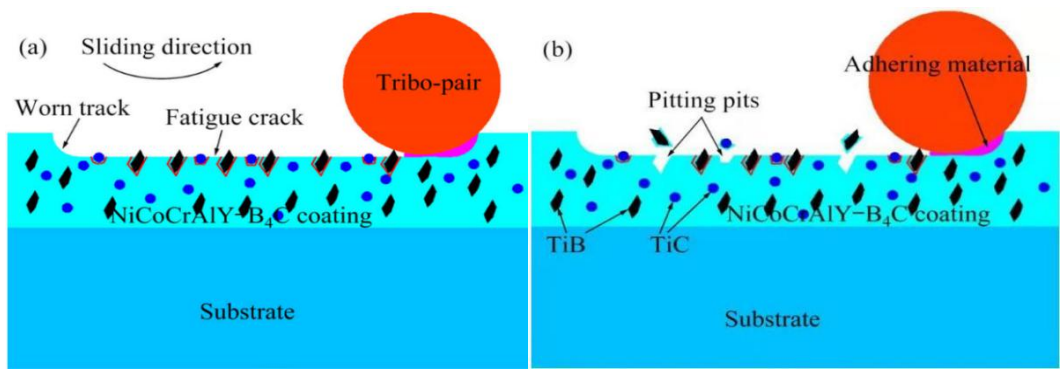

**Fig.6.** Sketches of wear mechanism of NiCoCrAlY-B$_4$C coatings in early (a) and normal (b) friction processes[55].

Hou [56] studied the influence of different content of B$_4$C on coating properties by using nickel-based alloy as raw material and B$_4$C powder as reinforcement phase. It was found that the hardness and wear resistance of the coating increased firstly and then decreased with the increasing of the content of B$_4$C hard phase in Ni60 alloy powder. When B$_4$C content (mass fraction) was 2%, the surface quality, hardness and wear resistance of the coating were the best. Wang et al. [55] used vacuum brazing to gain MCrAlY-$Cr_3C_2$ composite coating with a microhardness nearly 3 times higher than that of single crystal matrix. However, the distribution of CrC is not uniform overall. Due to the weakening of solid solution strengthening at the interface, the precipitation of hard and brittle phases has little effect on hardness, so the improvement of coating hardness is not obvious.

The above research results show that although a series of studies on carbide/nickel-based heterogeneous brazing coating have been carried out at home and abroad, most of them still remain in the laboratory stage. The internal mechanism of brazing coating is not clear yet, and the related methods and processes are immature, and there is a certain gap between the standardized achievement transformation and industrial production.

*3.4. Carbide/silver base heterogeneous brazing coating*

The silver base filler metal, mainly composed of silver and copper solid solutions, has moderate melting temperature, good wettability and excellent joint filling ability, so it is widely used in the field of heterogeneous brazing [70-72]. The bionic wear-resisting composite coating can be prepared by wetting the surface of c-BN cemented carbide particles with liquid Ag-Cu-Ti brazing filler. When the brazing temperature is 950℃ and the holding time is 20min, the reliable connection between the silver base solder and c-BN particles and 45# steel can be realized, and the bionic wear-resisting composite prepared shows good shape, smooth surface and uniform particle distribution[59].

Cao et al. [58] synthesized TiB whisker and tungsten particles in situ through the diffusion reaction between active Ti and WB particles, which were randomly distributed in the brazing joints.Fig. 7 illustrates the microstructure of the joints produced at 870℃ for 10 min.

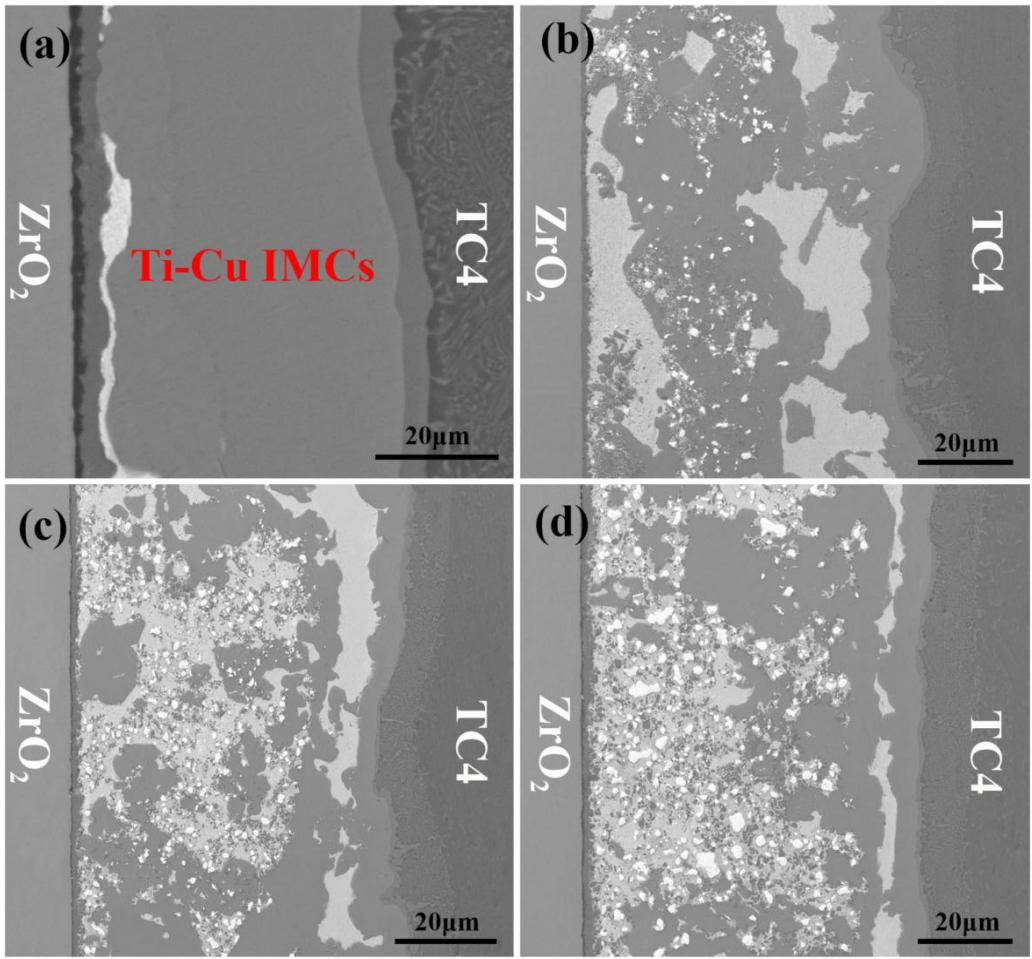

**Fig. 7.** Microstructure of the joints brazed at 870 °C for 10 min with different WB content: (a) 0 wt%; (b) 5 wt%; (c) 7.5 wt%; (d) 10 wt%[58].

The results have shown that adding WB particles in composite filler was an effective approach to improve the mechanical strength of the joints. During brazing, TiB whiskers and W particles were in situ synthesized in the brazing seam through the reaction of Ti and WB particles. Their formation served as nucleation sites for refining the microstructure of the joints and thus blocky Ti-Cu IMCs layers were dispersed in the brazing seam.

The active solder realized the highly reliable heterogeneous connection between diamond thick film and cemented carbide, which was prepared by sputtering titanium on the surface of Ag-Cu solder [57]. It was found that the continuous TiC film exists in the microstructure of brazing joint, which is generated by the reaction between active titanium and carbon atoms on diamond surface. Diamond film and brazing filler were connected by TiC film and active coating brazing filler by nailing.

## 4. The main application of cemented carbide heterogeneous brazing coating

*4.1. Life extension of soil-touching parts of agricultural machinery*

China is a typical agricultural country. Fast wear and failure of soil touching parts of agricultural machinery has always been an urgent problem to be solved. The surface wear-resistant coating is prepared based on brazing coating additive manufacturing. The additive material can combine two or more materials to form a functional body with special performance combination. The thickness of the additive coating is controllable, ranging from a few microns to a few millimeters or even thicker. It can also significantly reduce the wear ratio and effectively improve the hardness and fatigue strength. It is favored by the industry with high-quality cost performance.

Carbide composite brazing coating is effective in improving the service life of soil-touching parts of agricultural machinery. The author's team recently adopted diamond induction brazing technology in non-vacuum environment to prepare diamond brazing coating with 0.1~2.5 mm thickness on the surface of rotary blade[73]. As shown in Fig.8, the hardness of the coating is ⩾ 55 HRC. Compared with traditional tools, the wear resistance of the coating is increased by 5-8 times and the service life is increased by 4-10 times. The wear-resistant rotary tillage knife coating prepared by this method solves the problem of poor wear resistance of rotary tillage knife and greatly improves the service life of key components of agricultural machinery.

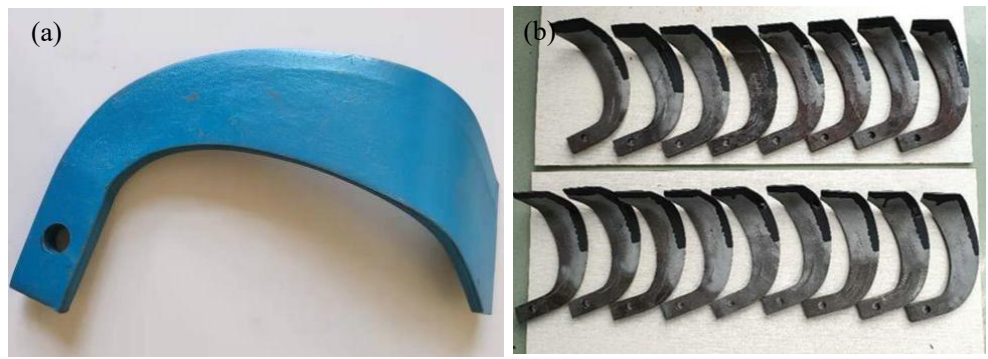

**Fig.8.** Pictures of rotating blades with diamond brazecoating
(a) Rotating blade; (b) surface topography of the brazecoating.

Xu et al. [34] developed WC/Cu non-smooth and high wear resistance composite coating with dung beetle surface characteristics by imitating non-smooth characteristics of biological surface. The hard carbide of the brazing coating supports load and resizes wear, while the soft matrix consolidates the hard alloy, which changes the scratching and chiseling of abrasive to rolling on the surface of the composite coating, effectively reducing the wear of abrasive to the composite coating. Compared with the traditional smooth plough-wall, the resistance reduction rate of the bionic non-smooth stick-reduction and drag reduction plough-wall [74] can reach

15%-18%, and the service life and wear resistance of agricultural machinery equipment can be increased by 30%-40%, which have opend up a new way to improve the wear resistance of easily worn parts such as soil touching parts of ground machinery. Ma et al. [51] carried out ploughing experiments on tools with different coating thicknesses, which showed that although coating thickness could improve the service life of the tool, it was not the thicker the better, only an appropriate thickness could make the tool self-sharpening better. The tool with a thickness of 60μm coating is much sharper with the best self-sharpening characteristic. The study provides a theoretical basis for the application of wear and self-sharpening ploughshare and stubble cutter in farmland. Due to the relatively short development time of heterogeneous brazing coating, its additive manufacturing mechanism, technology, special equipment and application effect need to be systematically studied.

*4.2. Repair of aviation components*

Being important components with complex structure and high cost, the working blade and guide blade of gas turbine components in aerospace field will endure pitting corrosion and crack after working for a certain period of time, which will seriously affect the stability and safety of aeroengine. Because of the high working temperature of the blade, only nickel base and cobalt base alloys with good heat resistance and high temperature strength can be used to repair the blade. However, most fusion welding processes destroy the structure of nickel-based and cobalt-based alloys. Therefore, brazing coating technology is the best solution for repairing damaged aviation components such as gas turbine components [75]. Xie et al.[76] investigated the influence of Cr addition on the interface purification of vacuum brazed NiCr-Cr3C2 coatings on single crystal superalloy, the results have shown that with the increase of the Cr addition, the coating/base metal interface was gradually purified(shown as Fig.9.).

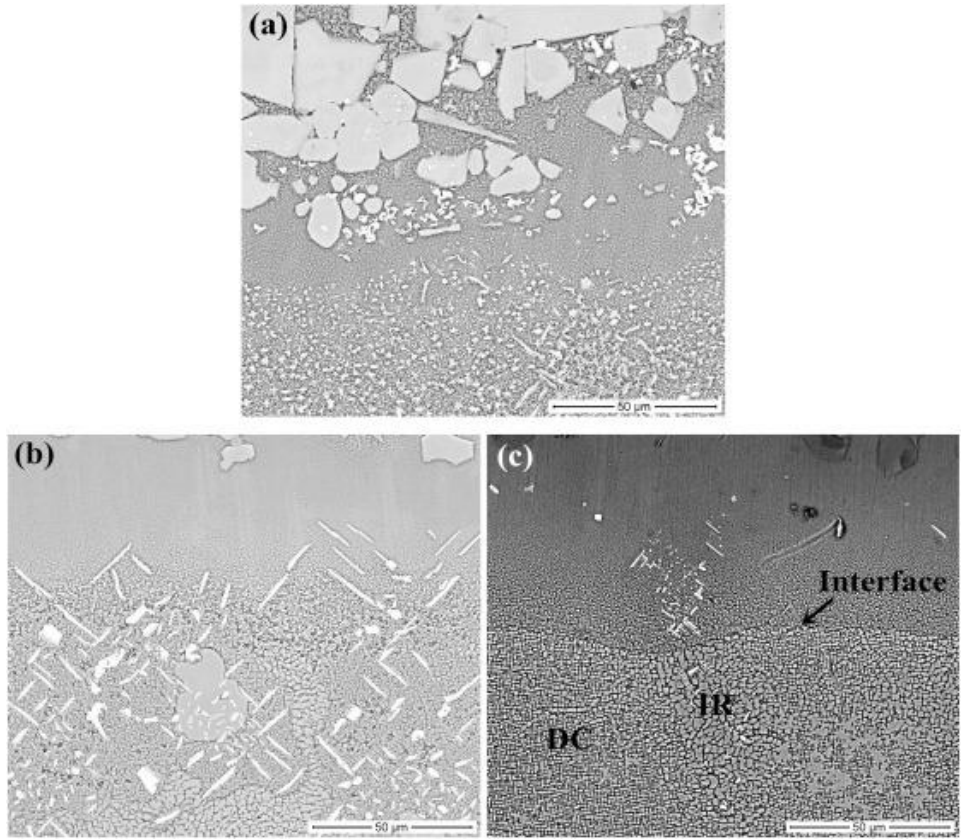

**Fig.9.** Cross sectional SEM images of interfaces for coatings with different Cr additions: (a)0 wt% Cr; (b)15 wt% Cr; (c)30 wt% Cr[76].

The application of pre-sintered sheet is an effective method for repairing damaged gas turbine components. Firstly, the powder similar to the base metal and the solder powder with low melting point are mixed, and then mixed with the organic binder to produce an adhesive sheet or adhesive band with a certain thickness, after that put the uniaxial pressure on the adhesive band to obtain high density adhesive band. Metal sheet would be processed by sintering adhesive tape in controlled atmosphere. Secondly, the pre-formed parts and turbine parts are welded together after assembly positioning, and then paste solder is added to realize the connection between matrix and coating by brazing. After sintering and heat treatment the pre-sintered sheet has a low porosity [75], so no shrinkage occurs in the brazing process. However, as a simple and reliable repair method, it is difficult for pre-sintered sheet with 3d complex shape to gain a good match with the matrix.

*4.3. Functional enhanced surface*

Wang [77] of East China University of Science and Technology found that the hardness and high temperature erosion resistance of the WC/Ni composite coating were greatly improved under vacuum brazing condition test which was carried out for the plane with abrasion failure. Based on the surface strengthening of radial bearing sleeve of screw drilling tool, Ba et al. [78] sampled and tested the interface. They found that the Ni60-WC composite coating prepared by vacuum cladding has a wear resistance 14 times higher than that of 45# steel, with tensile bond strength of 200MPa and surface microhardness as high as 950HV$_1$. In 3.5%NaCl solution with nitrogen deoxygenation, at 30℃, the self-corrosion potential of the composite coating is −588mV, which is suitable for general corrosion resistance environment.

Due to the strong bonding force between brazed WC coating and matrix and the small residual stress. Brazing coating technology has been used to prepare composite coating on the surface of electrical and thermal equipment components, which can achieve the purpose of wear resistance and corrosion resistance. Composite brazing coating can be designed on any surface with the help of the placement process, which has been widely applied on the surface of core components such as thermal equipment and fans of power plants in European and American developed countries [79]. La$_2$O$_3$ can effectively control the microstructure and properties of the brazing coating. After the addition of lanthanum oxide, the microstructure of WC-Co composite coating is much more uniform and dense, and the oxidation rate is greatly reduced [80-81]. This indicates that the oxidation resistance of WC brazing layer at high temperature can be improved by heating La$_2$O$_3$, which would broaden the application scope of composite brazing coating in the field of power and thermal equipment.

**5. Existing problems and prospect of cemented carbide heterogeneous brazing coating**

*5.1. Existing problems*

(1) Cemented carbide heterogeneous additive materials and technology need to be expanded

Because of the compatibility problem, high chromium and high boron iron-based wear-resistant materials, iron based, nickel based and cobalt based materials containing WC, TiC etc, chemical plating/permeability material are mainly additive materials in the applications of cemented carbide heterogeneous brazing coating at present, and the main direction of technology is arc additive manufacturing, laser additive manufacturing, plasma additive manufacturing and so on. Because different materials have different metallurgical and thermal properties, when these properties are incompatible, the bonding strength of the interface formed by them is weak, and defects such as cracking and delamination are easy to occur. Based on these problems, the interface of cemented carbide heterogeneous brazing coating additive parts is often the weakness of their mechanical properties, which will have an adverse impact on the overall performance of heterogeneous material parts and limit the application of cemented carbide heterogeneous brazing coating parts in many fields. In order to obtain high-quality brazing coatings, brazing coating additive materials and forming processes must be improved and expanded.

(2) The structure of cemented carbide heterogeneous brazing coating needs to be optimized

Due to the different movement modes and working environment of mechanical parts, the wear conditions are quite different. The brazing coating designed for one part can not be directly transplanted to other parts, and even the application effect of the same brazing coating is different in different service environments. At present, the application research mostly focuses on the preparation of composite coatings with wear-resistant and antifriction functions by using some relatively mature process, such as the preparation of wear-resistant layer on the surface of cemented carbide agricultural machinery cutters by laser cladding technology or thermal spraying technology, but the research on the structure design of cemented carbide brazed wear-resistant layer is insufficient, which leads to the unsatisfactory practical application.

(3) The mechanism of interface defects in cemented carbide heterogeneous brazing coating is unclear

Cemented carbide heterogeneous brazing coating additive manufacturing technology has great potential in the manufacture of heterogeneous material parts, which can give full play to the performance advantages of heterogeneous material parts and meet the urgent needs of modern society for high-performance parts. However, there are still many challenges in the process of exploiting the potential of cemented carbide heterogeneous brazing coating additive manufacturing technology. The precise presetting, data processing and control software of brazing coating additive materials are the key problems to be solved. In terms of interface forming and interface properties, the mechanism of interface defects in cemented carbide heterogeneous brazing coating is still unclear. How to effectively solve the performance problems caused by material incompatibility is also an important research direction.

*5.2. Expectation*

Although cemented carbide heterogeneous brazing coating additive manufacturing technology has incomparable advantages over traditional manufacturing methods and has shown great industrial value, it is still immature and has many deficiencies, so it has the following development trends:

(1) Interface properties and process control of brazing coating. It is very necessary to control the interface performance of heterogeneous brazing coating additive manufacturing parts for its wide application. On the one hand, the solution of material compatibility can be found through the expansion of additive materials, and the heterogeneous materials with compatible thermal and physical parameters can be selected to form the cemented carbide heterogeneous brazing coating parts, which is beneficial to obtain the heterogeneous brazing coating parts with good interface properties. On the other hand, excellent interfacial properties can be obtained by appropriate process control.

(2) Strengthening the structural design of cemented carbide brazing coating. Cemented carbide heterogeneous material additive manufacturing technology will tend to be integrated and intelligent in design and manufacturing. With the help of computer aided technology, the integrated design of material and structure can be carried out from the functional requirements, and the 3d model of cemented carbide heterogeneous brazing coating material parts could be obtained. After data processing, the 3D model can be uploaded to the forming equipment for parts manufacturing. The integrated design and manufacturing, intelligent manufacturing process is more in line with the actual demand, with higher efficiency and economic effect.

(3) Digital and intelligent control of the preparation process of brazing coating. Through the research on the interface defects of brazing coating in the manufacturing process of cemented carbide heterogeneous brazing coating additive, explore the scientific problems of material metallurgy behind it, and deeply understand the basic theory of heterogeneous material parts forming, so as to obtain high-performance heterogeneous brazing parts that meet the social needs. In order to improve the reliability of cemented carbide heterogeneous brazing coating additive manufacturing technology in actual production, the principle of physical compatibility and matching of cemented carbide heterogeneous materials need to be summarized, and the relationship database between fusion metallurgical parameters, thermal properties and interface forming quality between heterogeneous materials will be established, so as to realize the digital material combination and intelligent control of the manufacturing technology.

**Credit authorship contribution statement**

All the authors contributed equally to the manuscript.

**Declaration of competing interest**

The authors declare that they have no known competing financial interests or personal relationships that could have appeared to influence the work reported in this paper.

**Acknowledgements**

This work is supported by Henan Province's Major Key Technology Demand Unveiling and Tackling Key Projects (Grant No.191110111000), the State Key Laboratory of Advanced Brazing Filler Metals and Technology Projects(SKLABFMT202001).